\documentclass[preprint,12pt,authoryear]{elsarticle}

\usepackage{amssymb}
\usepackage{amsmath}
\usepackage{booktabs}  
\usepackage{caption}   
\usepackage{float}
\usepackage{array}   
\usepackage{rotating}
\newcommand{\Fig}[1]{Figure~\ref{#1}}
\newcommand{\Tab}[1]{Table~\ref{#1}}

\usepackage{flafter}        
\usepackage[section]{placeins} 
\usepackage{dblfloatfix}    

\usepackage{xcolor}
\definecolor{zc}{RGB}{0,100,0}
\newcommand{\zc}{\textcolor{zc}}

\makeatletter
\def\fps@figure{!htbp}
\def\fps@table{!htbp}
\makeatother

\journal{International Journal of Human-Computer Studies}

\begin{document}

\begin{frontmatter}
\title{Does Motion Intensity Impair Cognition in HCI? The Critical Role of Physical Motion-Visual Target Directional Congruency} 


\author[aff1,aff2]{Jianshu Wang} 
\author[aff5]{Siyu Liu}
\author[aff1,cor1]{Chao Zhou}
\author[aff1]{Yawen Zheng}
\author[aff1]{Hao Zhang}
\author{Yuan Yue}
\author[aff4]{Tangjun Qu}
\author[aff1]{Yang Li}
\author[aff2]{Yutao Xie}
\author[aff1]{Jin Huang}
\author[aff3,cor2]{Yulong Bian}
\author[aff1]{Feng Tian}

\cortext[cor1]{Corresponding author.}
\ead{zhouchao@iscas.ac.cn}

\cortext[cor2]{Corresponding author.}
\ead{bianyulong@sdu.edu.cn}

\affiliation[aff1]{organization={Institute of Software Chinese Academy of Sciences},
            addressline={}, 
            city={Beijing},
            postcode={}, 
            state={},
            country={China}}
            
\affiliation[aff2]{organization={Henan University},
            addressline={}, 
            city={},
            postcode={}, 
            state={},
            country={}}
            
\affiliation[aff3]{organization={Shandong University},
            addressline={}, 
            city={},
            postcode={}, 
            state={},
            country={}}    
\affiliation[aff4]{organization={Yanbian University},
            addressline={}, 
            city={},
            postcode={}, 
            state={},
            country={}}        
            
\affiliation[aff5]{organization={University College London},
            addressline={}, 
            city={},
            postcode={}, 
            state={},
            country={}} 
\begin{abstract}
Human-computer interaction (HCI) increasingly occurs in motion-rich environments. The ability to accurately and rapidly respond to directional visual cues is critical in these contexts. How whole-body motion \zc{and individual differences} affect human perception and reaction to these directional cues is therefore a key, yet an underexplored question for HCI. This study used a 6-DOF motion platform to measure task performance on a visual direction judgment task. We analyzed performance by decomposing the complex motion into two distinct components: a task-irrelevant lateral interference component and a task-aligned directional congruency component. Results indicate that increased motion intensity lengthened reaction times. This effect was primarily driven by the lateral interference component, and this detrimental impact was disproportionately amplified for individuals with high motion sickness susceptibility. Conversely, directional congruency, where motion direction matched the visual cue, improved performance for all participants. These findings suggest that motion's impact on cognition is not monolithic, and that system design for mobile HCI can be informed by strategies that actively shape motion, such as minimizing lateral interference while maximizing directional congruency.
\end{abstract}

\begin{graphicalabstract}
\begin{figure}[!h] 
    \centering 
    \includegraphics[width=\textwidth]{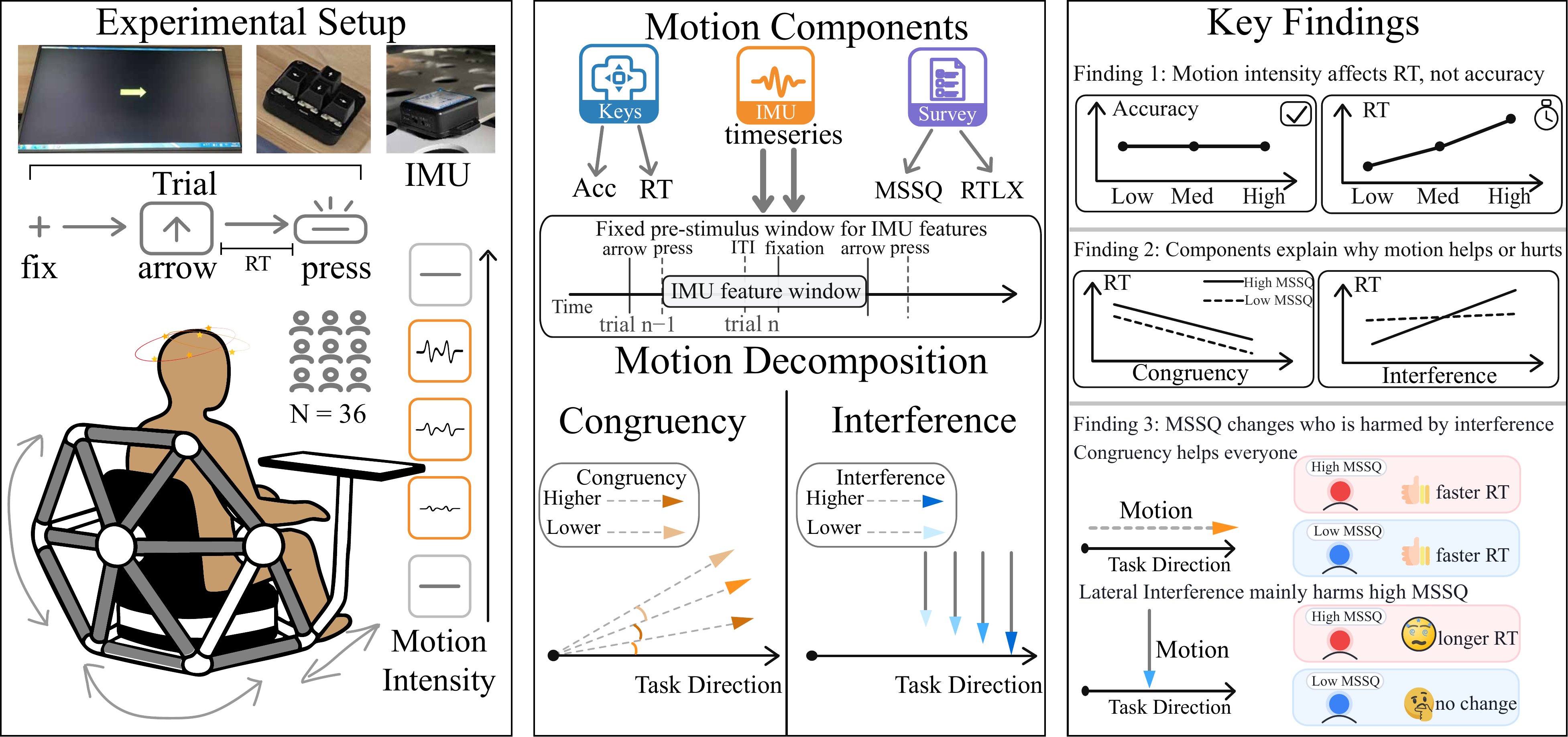} 
    \label{fig:graphicalAB}
\end{figure}
\end{graphicalabstract}

\begin{highlights}
\item Motion intensity lengthens reaction time, especially for highly susceptible individuals.
\item Directionally congruent motion consistently speeds reaction time across participants.
\item Lateral motion interference reliably slows reaction time and dominates the detrimental pathway.
\item Individuals' motion susceptibility selectively amplifies lateral-interference costs, not congruency benefits.
\item The results support motion shaping: minimize lateral interference while leveraging congruent cues.
\end{highlights}

\begin{keyword}
Motion Intensity \sep Motion-Visual Target Directional Congruency \sep Motion Sickness \sep Cognitive Performance \sep Individual Differences


\end{keyword}

\end{frontmatter}



\section{Introduction}

The advancement of autonomous driving is transforming vehicles from transportation tools into mobile environments for work, leisure, and social activities \citep{Berger2023Empowering}. This evolution prompts a fundamental shift in Human-Computer Interaction (HCI) design, moving from a driver-centric focus to a passenger-centric one where the quality of the user experience is paramount \citep{Ulm2022Acceptance}. Within this new paradigm, a fundamental and safety-critical task is the rapid and accurate judgment of directional visual cues, such as navigation arrows, warning signals, or takeover requests \citep{LU2016183, 10.1145/3409120.3410666}. The consequences of degraded performance on this task are significant \citep{LU2016183}. The defining challenge of this environment, however, is the vehicle's ever-present physical motion. This motion is not a benign background factor; it acts as an environmental stressor that degrades the user's cognitive state. Research shows that visual-vestibular conflict, a common occurrence when viewing a stable screen in a moving vehicle, is a primary source of cognitive fatigue \citep{10.1145/3654777.3676439}. Concurrently, whole-body vibration (WBV) transmitted through the seat has a causal relationship with physiological drowsiness \citep{BHUIYAN2022175}. This creates a challenging condition where the cognitive resources required to manage these motion-induced stressors directly compete with those needed to perform the primary directional judgment task. This gap motivates examining how vehicle motion shapes cognitive readiness, especially when current behavior in automated vehicles often falls short of the productive mobile office vision \citep{10.1145/3409120.3410662}.

Yet recent empirical evidence suggests that this mobile-office scenario is constrained by carsickness: a large-scale international survey reported that carsickness still affects about two-thirds of passengers and identified visually demanding in-transit activities as a major modulating factor \citep{Schmidt2020-carsicknessSurvey}. In open-road studies that systematically manipulated NDRTs, tasks with visual input---especially visually dynamic tasks---induced substantially more carsickness than tasks relying on internal vehicle displays alone \citep{Metzulat2024-visualInput}. Even basic interface choices such as vertical display placement (and thus the available peripheral outside view) can measurably influence carsickness outcomes \citep{Kuiper2018-displayPosition}.

To understand how physical motion impacts these directional judgments, we must turn to the underlying theoretical principles. Frameworks of embodied cognition posit that cognitive processes are not detached from the body, but are fundamentally grounded in its sensory and motor systems. From this perspective, physical motion is not background noise; it is a signal that must be processed. This processing can lead to two distinct, opposing outcomes. On one hand, task-irrelevant motion, particularly lateral motion, which has been shown to be uniquely demanding on attentional resources compared to other axes---acts as a potent source of sensory conflict \citep{10.1145/3654777.3676439}. This conflict requires the brain to expend active effort and cognitive resources to suppress the mismatch and maintain stability. This effort, in turn, directly competes with the resources needed for the primary visual judgment task. On the other hand, the principle of sensorimotor synergy suggests that motion is not always detrimental. If the physical motion is spatially or semantically aligned with the visual cue, it may act as a facilitative signal that enhances perception and performance, a principle demonstrated in related haptic and compatibility paradigms \citep{10.3389/frvir.2023.973083, Glenberg2002}. Therefore, the theoretical literature predicts that motion is a dual-edged sword, presenting a potential source of both detrimental interference and beneficial facilitation.

This theoretical duality, however, has rarely been investigated simultaneously within a single HCI context. To systematically isolate and quantify these opposing effects of motion, we designed an experiment using a 6-degrees-of-freedom (6-DOF) motion platform.
Participants were seated in the simulator and performed a visual direction judgment task, responding to arrows while exposed to different intensities of whole-body motion. We measured task performance, specifically reaction time and accuracy, and collected participants' baseline motion sickness susceptibility scores (MSSQ). The core of our analysis moved beyond comparing discrete intensity levels. We applied a functional geometric decomposition to the trial-level motion data recorded by sensors. This method allowed us to extract and analyze two key predictive components: a task-irrelevant Lateral Interference component and a task-aligned Directional Congruency component.
Our results provide empirical support for this dual-effect framework and reveal a key asymmetry. The analysis confirmed that the Lateral Interference component consistently predicted slower reaction times, confirming its detrimental effect. Conversely, the Directional Congruency component, which occurs when the physical motion matches the visual cue, predicted faster reaction times. Most importantly, we identified an asymmetric moderation by individual susceptibility. The detrimental impact of Lateral Interference was significantly amplified for individuals with high susceptibility. The facilitative benefit of Directional Congruency, however, was observed consistently across all participants, regardless of their susceptibility level. These findings provide empirical support for a more nuanced approach to HCI design in mobile environments, suggesting a shift from purely suppressing motion to actively shaping it, for instance by designing systems to minimize lateral conflict while leveraging directional synergy.

\section{Related Work}

Recent automated-vehicle research has made this constraint explicit: carsickness remains common in the general population and is repeatedly linked to how passengers allocate their visual attention to screens and other non-driving tasks \citep{Schmidt2020-carsicknessSurvey,Metzulat2024-visualInput}. From an HCI standpoint, this means that motion is not only a biomechanical disturbance but also a practical bottleneck for in-vehicle interaction time, task choice, and interface layout decisions such as display positioning \citep{Kuiper2018-displayPosition}. Seating orientation itself can also modulate sickness and user experience outcomes in automated-vehicle cabins, with rearward-facing arrangements reported to increase motion sickness and reduce trust during social non-driving tasks \citep{Rottmann2025-rearwardSeatingTrust}.

\subsection{Motion as an Environmental Stressor in HCI}

Alongside cue designs that aim to reduce motion sickness by increasing motion anticipation, empirical studies have tested peripheral visual feedforward displays and anticipatory auditory announcements of upcoming motion \citep{Karjanto2018-pvfs,KUIPER2020103068,Reuten2024-anticipatoryCues}. Importantly, the evidence is mixed: augmented visual environments can show ceiling effects when an external view already provides strong motion information \citep{deWinkel2021-augmentedVisual}, and increasing motion predictability does not guarantee reduced sickness if users do not perceive or exploit that predictability. This mixed pattern motivates a more task-grounded account of when motion becomes a resource rather than an uncontrollable stressor, which is precisely the question addressed by our motion decomposition.

In dynamic HCI environments, such as vehicles, physical motion is a primary factor influencing user performance. A large body of human factors research has established that physical motion, particularly Whole-Body Vibration, acts as an environmental stressor. Maintaining body and head stability against external perturbations is an active control process that requires attentional resources \citep{Shumway-Cook2000}. This occupies finite cognitive resources that are otherwise available to a primary task \citep{Shumway-Cook2000}. Furthermore, motion can induce sensory conflict, most notably the visual-vestibular conflict that occurs when visually perceived motion mismatches the motion sensed by the vestibular system \citep{ReasonBrand1975}. This conflict is not just a source of discomfort; it is a primary driver of cognitive fatigue \citep{Kourtesis2024Cybersickness, 10.1145/3654777.3676439}. Converging evidence has established a causal link between WBV and increased cognitive load, physiological drowsiness, and degraded cognitive task performance, such as slower reaction times \citep{BHUIYAN2022175, Jalilian2021}. This perspective establishes a foundational understanding of motion as a source of interference that must be managed in system design.

\subsection{Directional Cues as a Facilitator in Automotive HCI}

While physical motion often acts as an interferer, a separate line of research suggests that spatially-informative cues can act as a facilitator. When a physical motion or haptic cue is spatially or semantically aligned with the action required by a cognitive task, it can enhance performance \citep{Glenberg2002}. This principle has shown clear benefits in automotive HCI. A recent survey of haptic assistive driving systems concludes that spatially-informative vibrotactile or torque cues (e.g., on the steering wheel, seat, or belt) can shorten reaction times and reduce visual demand compared with purely auditory or visual warnings \citep{noubissie2022-hapticreview}. At the experiment level, classic studies demonstrate that spatially-congruent vibrotactile cues facilitate drivers’ visual attention and speed responses to time-critical events \citep{ho2005-trf}. In lane-keeping contexts, lane-departure warnings that encode direction haptically (e.g., belt or seat vibrations on the drifting side) elicit faster and smoother corrections than auditory beeps alone \citep{stanley2006-ldw}. This principle of multisensory integration, where the brain fuses information from different sensory channels to form a robust percept \citep{Kourtesis2024Cybersickness}, is also powerfully exemplified by pseudo-haptic feedback, where purely visual manipulations create compelling tactile illusions \citep{10.3389/frvir.2023.973083}. Together, these findings ground a key claim: when the geometry of physical motion (or its haptic proxy) is aligned with a demanded action, sensorimotor synergy can enhance performance.

\subsection{Embodied Cognition and Motion's Duality}
To understand the dual roles of motion identified in the previous sections, we adopt the perspective of Embodied Cognition \citep{Wilson2002}. This framework posits that cognitive processes are not abstract computations detached from the body, but are fundamentally grounded in the body's sensory and motor systems \citep{Wilson2002}. In HCI, this perspective has evolved into the paradigm of Embodied Interaction, which argues that interaction is not an abstract information processing task, but a skilled, engaged practice rooted in our physical and social world \citep{10.7551/mitpress/7221.001.0001}. Within this framework, physical motion is not merely background noise to be filtered out; it is an integral part of the dynamic sensorimotor context in which cognition occurs. The body's physical state is part of the essential context in which a visual judgment is made and constructed \citep{10.7551/mitpress/7221.001.0001}. This perspective provides a unified theoretical basis, explaining why motion can act as both an interferer (when it conflicts with the task) and a facilitator (when it is congruent with the task).

\subsection{Individual Differences: The Role of Susceptibility}
While the geometric relationship between motion and a cognitive task determines its potential to either interfere or facilitate, the ultimate cognitive outcome is not uniform across all individuals. The extent to which an individual is affected by these opposing forces of motion is highly variable. Motion Sickness Susceptibility (MSS) is known to be a robust predictor of an individual's response \citep{Golding2006}. The theoretical basis for this variability lies in the efficiency of a neural mechanism known as Sensory Re-weighting.

Sensory re-weighting is an adaptive process where the nervous system dynamically alters the weight, or importance, it assigns to different sensory channels based on their relative reliability in a given context \citep{10.1371/journal.pone.0260863}. When faced with a sensory conflict, the brain's first line of defense is to down-weight the unreliable sensory channel to maintain a stable percept. However, the ability to effectively re-weight sensory inputs varies across individuals. Recent research has established a direct link between deficits in sensory re-weighting capabilities and higher motion sickness susceptibility \citep{10.1371/journal.pone.0260863}. This suggests that an individual's susceptibility may be a behavioral marker for the underlying efficiency of this fundamental neural process. An individual with a less efficient re-weighting mechanism may be unable to effectively suppress salient but task-irrelevant motion cues, leading to greater interference with the primary visual task. This framework provides a compelling explanation for why the negative impact of sensory conflict is amplified in highly susceptible individuals.

\section{Experiment}
This study was designed to answer five research questions derived from our theoretical framework. RQ1 asked whether increasing motion intensity led to longer reaction time. RQ2 asked whether the magnitude of lateral interference predicted longer reaction time. RQ3 asked whether directional congruency predicted shorter reaction time. RQ4 asked whether motion-sickness susceptibility amplified the detrimental effect of lateral interference on reaction time. RQ5 asked whether this moderation was asymmetric, such that susceptibility primarily amplified the detrimental effect of lateral interference while having a smaller or negligible effect on the benefit of directional congruency.


\subsection{Participants}
A total of 40 participants (\zc{age= $22.3\pm 2.23$ years}
) were recruited for this study, including 22 males (\zc{age= $22.0\pm 2.35$ years} 
) and 18 females (\zc{age= $22.7\pm 2.08$ years} 
). \zc{All participants were right-handed with normal or corrected vision. There are no mental illnesses or vestibular related diseases.} Due to an error in data transmission, behavioral data from one participant were lost. \zc{The final valid data from 39 individuals would be included in subsequent analysis.} The study was carried out according to the Declaration of Helsinki guidelines and was approved by the Human Research Ethics Committee of our institution. Informed consent was obtained from each participant before the experiment started. After the experiment, the participants were briefed about the purpose of the study and received compensation for their participation.

\subsection{Experimental Design} 
\label{Experimetal Design}
The study employed a within-subjects design. The primary independent variable was motion intensity with five phases. The session started with a stationary baseline phase to establish each participant's performance before motion exposure. It then included three motion phases with low, medium, and high intensity, and the order of these three phases was counterbalanced across participants using a Latin-square design to control for order effects. The session ended with a stationary post-motion phase that was interpreted descriptively to check whether any behavioral or subjective changes persisted after motion ceased. The dependent variables were task accuracy and reaction time(RT), on the keypress task, together with NASA--TLX scores collected after each phase.

\subsection{Measures}
We measured three categories of variables: individual differences, behavioral performance, and subjective workload.

\subsubsection{Motion Sickness Susceptibility}
Upon arrival at the laboratory, participants completed the Motion Sickness Susceptibility Questionnaire. The total score was used in our analysis as a continuous predictor to account for individual differences in motion susceptibility.

\subsubsection{Behavioral Performance}
The primary dependent variables for the cognitive task were the participants' accuracy and reaction time on the keypress task.

\subsubsection{Subjective Workload}
To assess subjective workload, participants were required to fill out the NASA-TLX questionnaire immediately after the completion of each experimental block. Our analysis focused on the Raw TLX total score (RTLX Score) as the indicator of overall perceived workload.

\subsection{Apparatus and Materials}
The experiment was conducted on a 6-DOF electric motion platform. Participants were seated in a racing seat equipped with a four-point safety harness. A 16.1-inch display with 2.5K resolution and a 180 Hz refresh rate was positioned horizontally in front of the participant. Two inertial measurement units (IMUs) were placed beneath the seat to record the veridical acceleration data of the motion platform during motion.

\begin{figure}[!t] 
    \centering 
    \includegraphics[width=\textwidth]{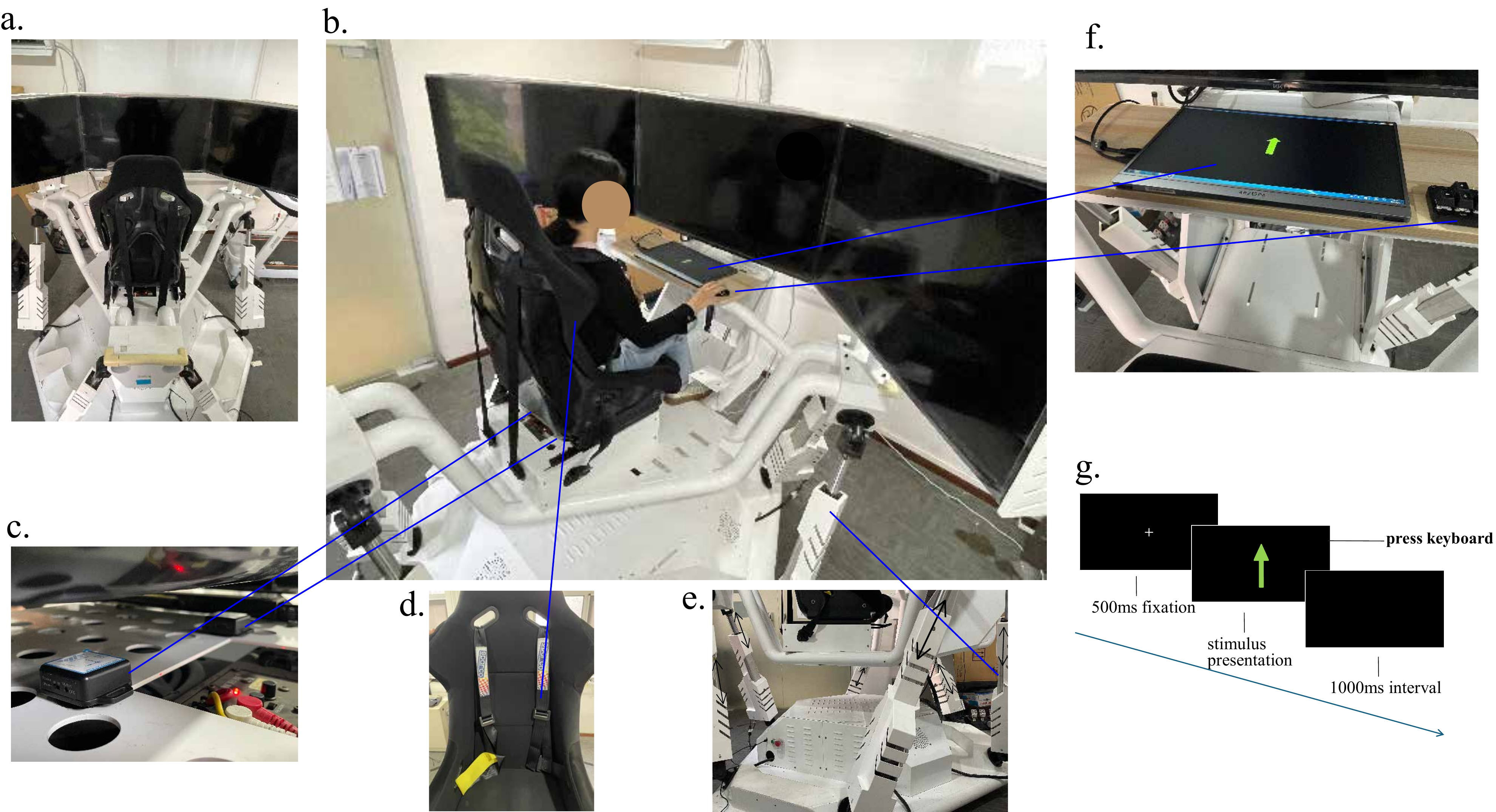} 
      \caption{Experimental setup and procedure. (a--b) The 6-DOF motion platform and participant seating. (c--e) Hardware details showing IMU placement, safety harness, and actuators. (f) The visual direction judgment task interface. (g) Timeline of a single trial.}
    \label{fig:apparatus} 
\end{figure}

\subsection{Task and Stimuli}

\subsubsection{Task Paradigm}
The task paradigm can be seen in \Fig{fig:apparatus}g. \zc{The stimulus was presented in the center of the screen with a black background.} A single trial began with a 500~ms fixation cross presented at the center of the screen. Following the fixation, an arrow was presented, and the participant's task was to press the corresponding direction key as accurately and quickly as possible. The stimulus was followed by a 1000~ms blank screen as inter-trial interval. There were four arrow directions (front, back, left, right), each appeared 15 times per block (60 trials per block in total). Each participant completed five blocks for a total of 300 trials.

\subsubsection{Motion Stimuli Generation}
We selected three levels of motion intensity based on the frequency-weighted root-mean-square acceleration $a_{w}$. The target values were 0.1~m/s$^{2}$ for the low condition, 0.2~m/s$^{2}$ for the medium condition, and 0.4~m/s$^{2}$ for the high condition.

The foundation for this approach is the ISO 2631-1 standard on human exposure to whole-body vibration \citep{ISO2631-1}. This standard defines vibration levels below $0.315~m/s^{2}$ as ``not uncomfortable''. Our Low and Medium intensity conditions are set well within this comfort zone. Our High intensity condition at $0.4~m/s^{2}$ was strategically chosen to be at the edge of this comfort boundary. While this level enters the ``a little uncomfortable'' range in the ISO standard, it remains below the empirically-derived threshold where discomfort actually begins. Foundational research from NASA established this threshold at approximately $0.422~m/s^{2}$ \citep{Leatherwood1979}. By selecting these three levels, we created a spectrum of motion that is, by most engineering accounts, acceptable. The actual acceleration gradients achieved across the experimental conditions are illustrated in \Fig{fig:IMU}a, confirming the distinct intensity levels.

\subsubsection{Cognitive Motion Components}
To investigate the impact of physical motion on cognitive performance at a finer-grained level, we extracted two core cognitive motion components from the real-time three-axis acceleration data recorded by the IMU for each trial. These components were designed to decompose the complex 6-DOF motion into quantifiable predictors with distinct cognitive meanings within the context of the current task.

The calculation was based on a unified coordinate system aligned with the cockpit's motion, where the x-axis pointed forward, the y-axis pointed left, and the z-axis pointed upward. The motion characteristics for each trial were derived from the average 3D acceleration vector, $\vec{v}_{motion}$, calculated during the time window preceding the stimulus presentation (i.e., between the response of the previous trial and the stimulus onset of the current trial). We deliberately derived trial-level motion features from the interval immediately preceding stimulus onset, defined from the previous response to the current stimulus onset, rather than from the period between stimulus onset and the response. This choice preserves the temporal order needed for interpretation, because the motion estimate is defined before the behavioral response in the current trial is produced. The interval is largely set by the task timing, including the inter-trial interval and the fixation period, which provides a consistent definition across trials. It also avoids a key source of circularity: if the averaging window were determined by the current trial reaction time, the predictor would be partly shaped by the outcome it aims to explain. In practical terms, the resulting vector captures the ongoing body motion state at the moment the arrow appears, allowing us to test whether motion aligned with the required response direction supports faster responses, whereas motion orthogonal to that direction is associated with slower responses.

We define the directional congruency component to quantify the alignment between the net cockpit motion and the task-relevant axis. We converted the planar arrow directions into 3D unit vectors $\vec{v}_{arrow}$, for example forward as [1, 0, 0], backward as [-1, 0, 0], left as [0, 1, 0], and right as [0, -1, 0]. Directional congruency was calculated as the dot product between the arrow vector and the trial-level motion vector $\vec{v}_{motion}$.
\begin{equation}
\text{Directonal Congruency} = \vec{v}_{arrow} \cdot \vec{v}_{motion}
\end{equation}
The resulting scalar provides a continuous index of alignment weighted by motion magnitude. A large positive value indicates that cockpit motion reinforces the task direction, while a large negative value indicates that motion opposes the task direction.

We define the lateral interference component to isolate task-irrelevant motion in the horizontal plane that is orthogonal to the required response direction. We focus on this orthogonal horizontal component because evidence from dual-task paradigms suggests that managing mediolateral stability draws more on limited attentional resources than fore-aft control. The computation proceeded by projecting the 3D motion vector $\vec{v}_{motion}$ onto the horizontal plane to obtain $\vec{v}_{hor}$. We then projected $\vec{v}_{hor}$ onto the task direction vector $\vec{v}_{arrow}$ to obtain $\vec{v}_{proj}$, and computed the orthogonal residual as $\vec{v}_{side}=\vec{v}_{hor}-\vec{v}_{proj}$. Lateral interference was defined as the magnitude of this orthogonal vector.
\begin{equation}
\text{Lateral Interference} = ||\vec{v}_{side}||
\end{equation}
A larger value indicates stronger task-irrelevant physical interference.

Figure \ref{fig:IMU}c visually summarizes the trial-level distribution of these two extracted components. As shown in the hexbin plots, the experimental motion profiles successfully elicited a broad coverage of the parametric space, with distinct distributions of lateral interference and directional congruency across the intensity levels.

\subsection{Procedure}

Upon arrival at the laboratory, participants first read and signed an informed consent form, after which they completed  a basic demographic information questionnaire and MSSQ. Next, participants were seated in the motion simulator, where the experimenter provided detailed instructions for the task. Participants were required to complete a short practice block (approximately 2 minutes) to ensure full comprehension of the task and to reach a stable baseline performance.

\begin{figure*}[!t]
  \centering
  \includegraphics[width=\textwidth]{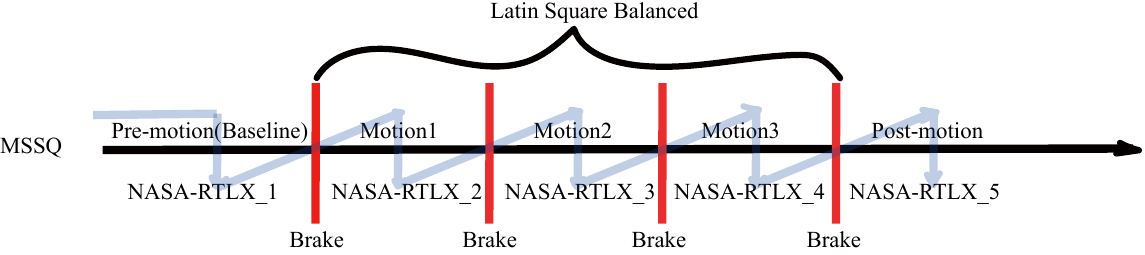}
  \caption{Session timeline. Participants completed a pre-motion baseline block, three motion blocks labelled Motion1 to Motion3 with the order counterbalanced using a Latin-square design, and a post-motion block. Braking intervals separated phases. NASA--TLX was administered after each phase and the raw total score was used as RTLX for analysis, spanning RTLX\_1 to RTLX\_5. MSSQ was assessed once per participant to quantify motion-sickness susceptibility.}
  \label{fig:procedure}
\end{figure*}

The formal experiment consisted of the five blocks described in \ref{Experimetal Design}. Each block consisted of 10 practice trials followed by 60 formal trials. 
After the completion of each block, participants were required to immediately fill out the NASA-TLX questionnaires. 
The entire experimental session, including preparation, practice, and formal experiment, lasted approximately 45 minutes. An overview of the procedure can be seen in \Fig{fig:procedure}.

\subsection{Statistical Analysis}


\zc{data preprocessing}



The analysis comprised two main parts: The first part examined the overall impact of different motion conditions on \zc{behavioural performance (}accuracy and reaction time\zc{)} using GLMM and LMM, respectively. The second part delved deeper to investigate how the motion components derived from IMU data influenced individual reaction times, using an LMM. Confirmatory analyses focused on the core motion blocks and their interactions with motion susceptibility in the mixed-effects models.

All statistical analyses were conducted in R 4.5.1 to systematically test our hypotheses. Our primary analytical strategy centered on using Linear Mixed-Effects Models (LMMs) to probe our core research questions regarding cognitive performance. To investigate our research question, we developed two key LMMs predicting reaction time: the first model was designed to examine the overall impact of Motion Intensity and its interaction with susceptibility (testing RQ1) , while the second, more granular model, was built to investigate the opposing effects of the decomposed Directional Congruency and Lateral Interference components and their asymmetric moderation by susceptibility (testing RQ2, RQ3, RQ4, and RQ5). To complement this primary analysis, a Generalized Linear Mixed-Effects Model (GLMM) was used to assess task accuracy. Finally, to understand the impact on subjective experience, we analyzed questionnaire data using the Aligned Rank Transform (ART) ANOVA, a non-parametric method suitable for ordinal and non-normally distributed data.

We chose LMMs and GLMMs over traditional repeated-measures ANOVA because they do not require the strict sphericity assumption to be met and can flexibly incorporate both categorical and continuous predictors within a unified model.

For the analysis of subjective questionnaire data, we first evaluated whether the scores met the normality assumption required for parametric tests. Visual inspection of the Q--Q plots showed marked deviations from normality for the NASA-TLX total score across \zc{motion} phases (see Appendix~A, Fig.~\ref{fig:nasa_qq}), which was further supported by Shapiro--Wilk tests (all $p < .05$). 
Given this violation, we analyzed all questionnaire outcomes using the ART ANOVA, a non-parametric method suitable for repeated-measures designs.

\subsubsection{Behavioral Performance Analysis}
To examine the impact of motion conditions on cognitive performance, we constructed a GLMM for accuracy and an LMM for RT. The analysis used unaggregated trial-level data. For RT, we included correct trials only and excluded responses shorter than 200~ms or longer than 2500~ms. We log-transformed RT to reduce positive skew and improve residual normality. Fixed effects included the five-level motion condition and motion-sickness susceptibility, with susceptibility centered and scaled. All models included a random intercept for participant ID to account for baseline individual differences. We selected the final fixed-effects structure using likelihood ratio tests that compared a full model containing the motion condition by susceptibility interaction against a reduced model containing main effects only, and models for these comparisons were fitted using maximum likelihood. Model diagnostics were performed using the performance package in R.

For the reaction time model (LMM), visual inspection of the diagnostic plots confirmed that all key assumptions were met. Specifically, the posterior predictive check showed an excellent fit between model-predicted and observed data. The residuals were normally distributed and showed no discernible patterns in the linearity and homogeneity of variance plots. Furthermore, the Variance Inflation Factors (VIFs) were well below 5, indicating no issues with multicollinearity. The random effects were also confirmed to be normally distributed. 

For the accuracy model (GLMM), a similar set of diagnostics was performed. The binned residuals plot and the posterior predictive check both indicated a good model fit. The check for uniformity of residuals confirmed a proper residual distribution. Similar to the LMM, there were no multicollinearity issues, and the random effects were normally distributed. While a few influential observations were identified, they did not compromise the overall validity and robustness of the model.

\subsubsection{User Experience Analysis}
For the NASA-TLX questionnaire data, we used ART ANOVA, implemented via the ARTool package \citep{elkin2021aligned}. This non-parametric method was chosen because subjective questionnaire scores are inherently ordinal data and their distributions often violate the normality assumption required by traditional parametric tests (see \Fig{fig:nasa_qq}). ART-ANOVA analyzes the ranked data, providing a statistically robust hypothesis testing method that does not assume normality, while still being able to handle the within-subjects design of our study by including a random effect for participant ID.

For the NASA-TLX data, our inferential statistical analysis focused on the Raw TLX total score (RTLX Score) to test the effect of motion conditions on overall cognitive load. For all significant main effects, we conducted post hoc pairwise comparisons using the art.con function.

\subsubsection{Motion Component Analysis}
To investigate the impact of specific motion patterns, we constructed a third LMM to predict RT. The dependent variable was the filtered and log-transformed RT. Fixed effects included directional congruency, lateral interference, motion-sickness susceptibility, and their two-way interactions, and all continuous predictors were mean-centered. The model included a random intercept for participant ID. We selected the final model using likelihood ratio tests that compared a full model with all two-way interactions against a reduced model with main effects only, using maximum likelihood estimation for the comparison. For significant fixed effects in the final model, we conducted Tukey-adjusted post hoc analyses, and the alpha level was set to 0.05. We assessed residual normality, homoscedasticity, and influential observations using standard diagnostic plots and influence measures, and these checks supported the validity of the model assumptions.

First, the assumption of normality of residuals was assessed. The Q-Q plot of the model residuals showed that the points closely followed the diagonal line, indicating that the residuals were approximately normally distributed. Second, the homoscedasticity assumption was evaluated by plotting the standardized residuals against the fitted values. The scatter plot revealed no discernible patterns and a random distribution of points, suggesting that the residuals were homoscedastic and that the variance of the residuals was constant across the range of fitted values.

Furthermore, we examined for the presence of outliers and influential data points. No severe outliers with standardized residuals exceeding an absolute value of 3 were identified. The analysis of influential observations, based on Cook's Distance, showed that no single trial exerted a disproportionate influence on the model's coefficient estimates.

These diagnostic analyses confirm that the model assumptions were sufficiently met, thereby supporting the reliability and robustness of our findings.

\section{Results}

\subsection{Impact of Motion Intensity on Behavioral Performance}

\subsubsection{Accuracy}
The Likelihood Ratio Test comparison of the models revealed that the interaction effect was not significant, $\chi^2(2) = 3.86, p = .425$. Therefore, we selected the more parsimonious main effects model for the final report.

The main effect of motion condition was not significant, $\chi^2$(4) = 5.56, $p$ = .234. The main effect of motion sickness susceptibility was also not significant, $\chi^2$(1) = 1.39, $p$ = .238.

\subsubsection{Reaction Time}

\begin{figure}[!t] 
    \centering 
    \includegraphics[width=0.8\textwidth]{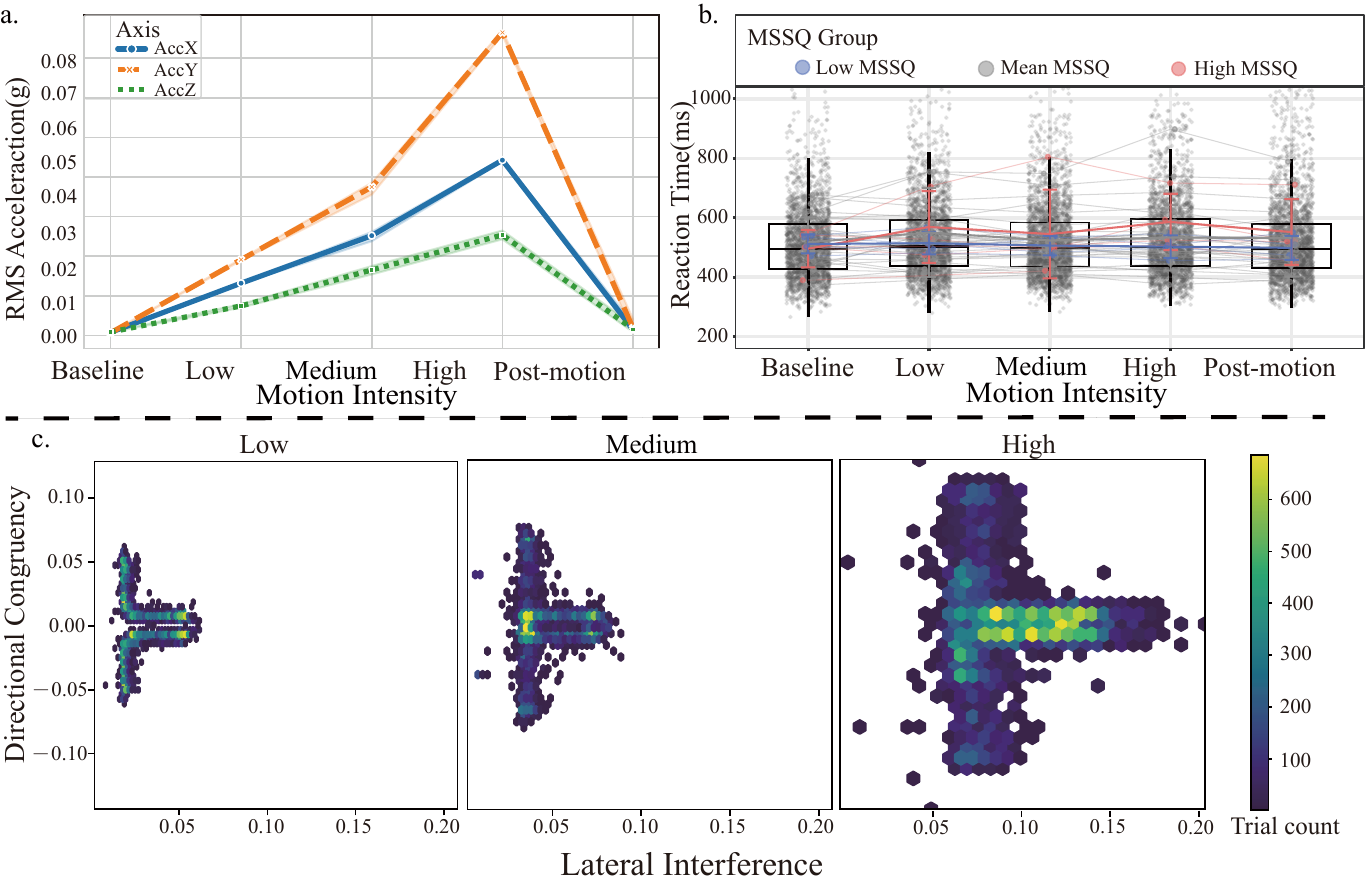} 
    \caption{Descriptive checks of platform motion and reaction time across the five motion conditions.
(a) Root-mean-square (RMS) acceleration (g) on each IMU axis (AccX/AccY/AccZ) as a function of motion condition (Baseline, Low, Medium, High, Post-motion).
(b) Trial-level reaction time (RT, ms) distributions across motion conditions for correct trials after standard RT filtering (gray points); boxplots summarize the trial-level distributions.
Thin lines connect each participant’s condition-wise mean RT to show within-participant trajectories; colored overlays summarize group-level trends by MSSQ susceptibility (Low/Mean/High).
(c) Hexbin density of the two motion components (Lateral Interference vs.\ Directional Congruency) for the three motion-present conditions (Low/Medium/High), with color indicating trial counts.}
    \label{fig:IMU} 
\end{figure}

The likelihood-ratio test favored the full model over the reduced model, $\chi^2(4) = 64.32, p < .001$. Therefore, the full model including the interaction term was selected for subsequent analysis. The final model demonstrated a good overall fit. Conditional $R^2$ was $.53$ and marginal $R^2$ was $.01$. The intraclass correlation coefficient was $ICC = .53$, indicating that a large share of variance reflected stable between-participant differences rather than within-participant condition-to-condition changes. Descriptively, \Fig{fig:IMU}a summarizes the RMS acceleration along the three IMU axes across the five phases, providing a manipulation check of the intended motion-intensity profile. \Fig{fig:IMU}b visualizes the corresponding trial-level RT distributions across phases (single-trial dots with boxplots), with model-based trends overlaid at the three MSSQ probe values.

\begin{figure*}[!t]
  \centering
  \includegraphics[width=\textwidth]{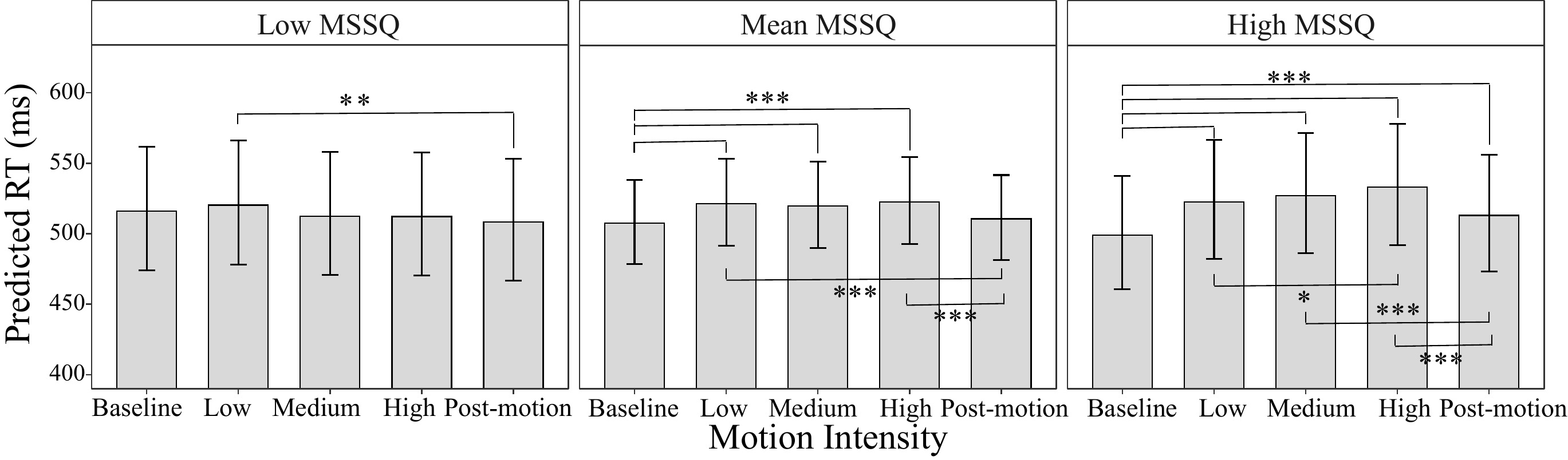}
  \caption{Post-hoc probe of the Motion Condition $\times$ MSSQ interaction for reaction time. Bars are model-estimated marginal means of back-transformed RT (ms) at different levels of MSSQ; error bars are 95\% Wald confidence intervals. Asterisks denote Tukey-adjusted pairwise contrasts within each MSSQ level ($^{*}p<.05$, $^{**}p<.01$, $^{***}p<.001$).}
  \label{fig:rt_posthoc_cld}
\end{figure*}

Given the significant Motion Condition $\times$ MSSQ interaction, we probed simple effects using estimated marginal means (EMMs) and Tukey-adjusted pairwise comparisons of motion conditions at three MSSQ probe values corresponding to $-1$, $0$, and $+1$ $\mathit{SD}$ of the scaled MSSQ score, as summarized in \Fig{fig:rt_posthoc_cld}. Contrasts are reported on the log(seconds) scale as $\hat{\beta}$, and the exponentiated effect is reported as a multiplicative change in reaction time with ratio $=\exp(\hat{\beta})$.

For low MSSQ, only Low intensity was slower than Post-motion (ratio $=1.02$, $p_{\mathrm{Tukey}}=.008$).

For mean MSSQ, Low, Medium, and High were slower than Baseline (Low: ratio $=1.03$, $p_{\mathrm{Tukey}}<.001$; Medium: ratio $=1.02$, $p_{\mathrm{Tukey}}<.001$; High: ratio $=1.03$, $p_{\mathrm{Tukey}}<.001$), and they remained slower than Post-motion (Low: ratio $=1.02$, $p_{\mathrm{Tukey}}<.001$; Medium: ratio $=1.02$, $p_{\mathrm{Tukey}}=.003$; High: ratio $=1.02$, $p_{\mathrm{Tukey}}<.001$).

For high MSSQ, motion-induced slowing was stronger: Low, Medium, and High were slower than Baseline (Low: ratio $=1.05$, $p_{\mathrm{Tukey}}<.001$; Medium: ratio $=1.06$, $p_{\mathrm{Tukey}}<.001$; High: ratio $=1.07$, $p_{\mathrm{Tukey}}<.001$), and Post-motion was also slower than Baseline (ratio $=1.028$, $p_{\mathrm{Tukey}}=.001$). Medium and High remained slower than Post-motion (Medium: ratio $=1.03$, $p_{\mathrm{Tukey}}=.001$; High: ratio $=1.04$, $p_{\mathrm{Tukey}}<.001$).

A detailed summary of all pairwise comparisons can be referred to \Tab{tab:posthoc_pairs_by_mssq}.

The ANOVA for the final model confirmed a significant interaction effect between motion condition and MSSQ score, $F(4, 12406.8) = 16.11, p < .001$.

\subsection{Impact of Motion Intensity on User Experience}
To assess the impact of different motion conditions on participants' subjective feelings, we conducted Aligned Rank Transform (ART) ANOVAs on the data from the NASA-TLX Task Load Index.

\begin{figure}[!t] 
    \centering 
    \includegraphics[width=0.7\textwidth]{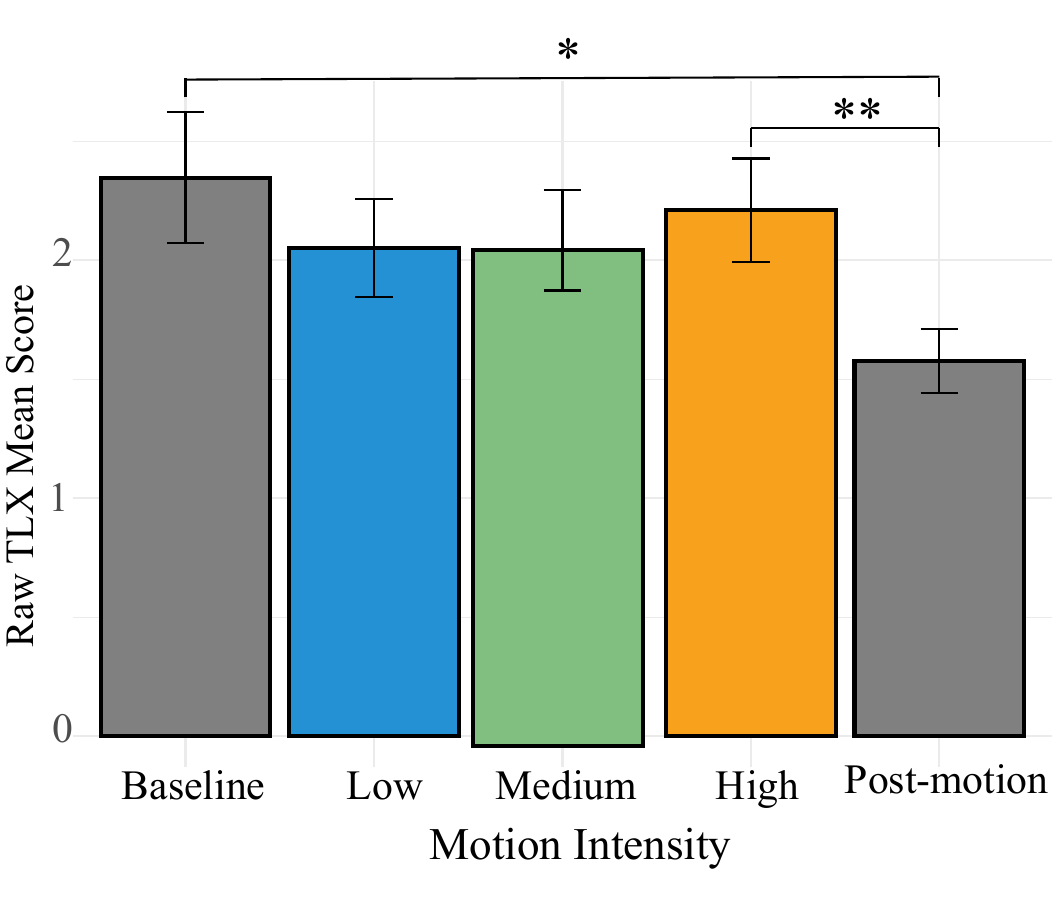} 
    \caption{Mean NASA-TLX (RTLX) total scores across motion conditions (Baseline, Low, Medium, High, Post-motion).
Error bars denote 95\% confidence intervals of participant means.
Asterisks indicate significant post-hoc pairwise differences (* $p<.05$, ** $p<.01$).}
    \label{fig:rtlx-bar} %
\end{figure}

The ART-ANOVA on the RTLX total scores revealed a significant main effect(see \Fig{fig:rtlx-bar}) of motion condition on the participants' perceived overall workload, $F_{(4, 161.13)}$ = 4.11, $p$ = .003, $\eta_p^2$ = .09.

However, post hoc pairwise comparisons revealed that this significance was primarily driven by changes during the final phase of the experiment. Specifically, the workload reported during the final stationary condition (post-motion, $M$ = 1.58, $SD$ = 0.85) was significantly lower than that reported during the initial stationary condition (baseline, $M$ = 2.35, $SD$ = 1.74; $p$ = .012) and also significantly lower than during the high-intensity motion condition (high, $M$ = 2.21, $SD$ = 1.37; $p$ = .003). The differences in workload between the motion conditions themselves and the initial baseline were not significant.

\subsection{Impact of Motion Components on Reaction Time}
The Likelihood Ratio Test result showed that the full model including interaction terms was significantly better than the reduced model with only main effects, $\chi^2(2) = 49.453, p < .001$. We therefore selected the full model for the final analysis. Conditional $R^2$ was $.546$ and marginal $R^2$ was $.018$. The intraclass correlation coefficient was $ICC = .538$, indicating substantial clustering of reaction time by participant. To contextualize the two derived predictors, \Fig{fig:IMU}c shows the joint distribution of Directional Congruency and Lateral Interference during the low-, medium-, and high-intensity motion phases.

\begin{figure}[!h] 
    \centering 
    \includegraphics[width=\textwidth]{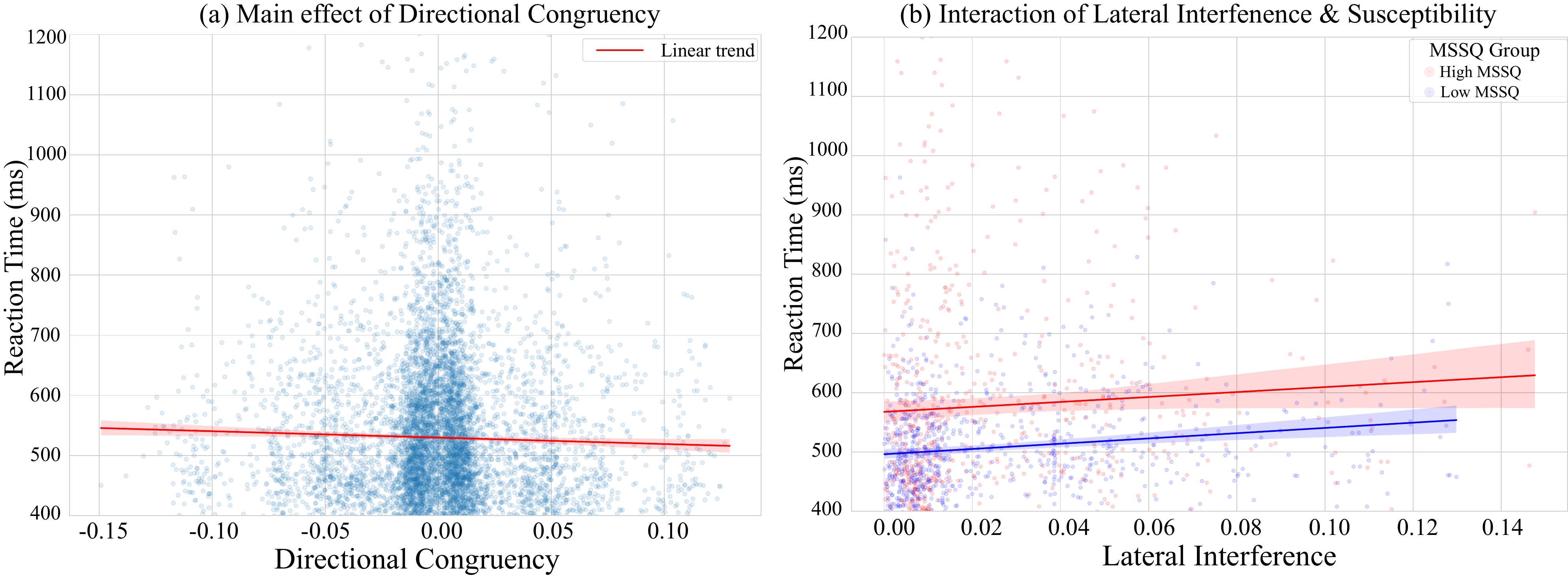} 
    \caption{Effects of motion components on reaction time. (a). The main effect of the directional congruency component. (b). The interaction effect between the lateral interference component and individual MSSQ. Curves show predictions at low/mean/high MSSQ; shaded bands indicate 95\% CIs; reaction time is in ms.}
    \label{fig:component_rt} 
\end{figure}

The detailed fixed-effects coefficients of the model are summarized in \Tab{tab:lmm-coeffs}. The model revealed a significant negative main effect of the congruent component ($\beta = -0.33, p < .001$) in \Fig{fig:component_rt}a, and a significant positive main effect of the lateral component ($\beta = 1.49, p < .001$) in \Fig{fig:component_rt}b. Crucially, the model identified a significant interaction between the lateral component and motion sickness susceptibility ($\beta = 0.018, p < .001$). In contrast, the interaction between the congruent component and susceptibility was not significant ($p = .179$).

\begin{table}
\centering
\caption{Fixed-effects estimates from the linear mixed-effects model predicting log-transformed reaction time (seconds) from the two motion components, MSSQ, and their interactions.
The table reports coefficient estimates, standard errors, degrees of freedom, $t$ statistics, and $p$ values.}
\label{tab:lmm-coeffs}
\begin{tabular}{lrrrrr}
\toprule
\textbf{Predictor} & \textbf{Estimate ($\beta$)} & \textbf{Std. Error} & \textbf{df} & \textbf{t value} & \textbf{p value} \\
\midrule
(Intercept) & -0.65 & 0.03 & 37.8 & -23.19 & $< .001$*** \\
\textit{Main Effects} & & & & & \\
\quad Congruent Comp. & -0.33 & 0.05 & 11974.7 & -6.74 & $< .001$*** \\
\quad Lateral Comp. & 1.49 & 0.09 & 11975.4 & 17.53 & $< .001$*** \\
\quad MSSQ Score & 0.00 & 0.002 & 36.6 & 0.04 & .970 \\
\textit{Interaction Effects} & & & & & \\
\quad Lateral Comp. $\times$ MSSQ & 0.02 & 0.003 & 11976.3 & 6.88 & $< .001$*** \\
\quad Congruent Comp. $\times$ MSSQ & -0.004 & 0.003 & 11974.7 & -1.34 & .179 \\
\bottomrule
\end{tabular}
\end{table}

To decompose the significant interaction between the lateral component and MSSQ, a simple slope analysis was conducted. The results showed that the detrimental effect of the lateral component on reaction time was amplified by susceptibility. For low susceptibility participants (MSSQ = -1 SD), the slope was significantly positive ($\beta = 1.18, p < .001$). For medium susceptibility participants (MSSQ = Mean), the slope became steeper ($\beta = 1.49, p < .001$). For high susceptibility participants (MSSQ = +1 SD), the slope reached its maximum value ($\beta = 1.80, p < .001$).

\section{Discussion}

A similar dissociation between subjective state and objective performance has been reported in recent test-track studies of automated driving: induced car sickness substantially reduced perceived fitness to drive and increased perceived criticality, yet emergency braking reaction times and slalom performance showed little or no objective degradation \citep{Metzulat2025-stillFitDrive}. This convergence strengthens the interpretation of hidden costs in sustained-motion interaction: users can maintain performance by reallocating effort and control resources, but that compensation is itself consequential for fatigue, comfort, and longer-duration interaction.

Beyond our decomposition approach, independent ergonomics work has shown that whole-body vibration can measurably slow reaction time and increase perceived workload in dual-task settings \citep{Jalilian2021}, and international standards formalize how vibration magnitude should be weighted across frequency when evaluating human exposure \citep{ISO2631-1}. This broader evidence base supports treating vibration characteristics as more than a nuisance variable when reasoning about mobile interaction performance.

This study was designed to supplement the prevailing view in HCI that treats physical motion as a monolithic nuisance to be suppressed. To achieve this, we proposed and implemented a functional geometric decomposition, using a high-fidelity motion simulator to isolate motion's opposing effects on cognition: a detrimental Lateral Interference and a beneficial Directional Congruency. Our findings fundamentally validate this approach, demonstrating that motion is, in fact, a double-edged sword with functionally distinct and opposing effects on cognition. Our results show that while increasing motion intensity imposes a hidden cognitive cost, primarily affecting individuals with high motion sickness susceptibility, this macroscopic effect is the result of these two competing microscopic forces. Most profoundly, our findings reveal a fundamental asymmetry in the human cognitive system's response to these forces: while the detrimental impact of interference is amplified by individual susceptibility, the performance benefit from directional synergy is universal, applying robustly across all individuals.

\subsection{The Hidden Cognitive Load and Compensatory Control}
Our first research question examined the overall impact of motion intensity on task performance. The results show that while participants' accuracy in the directional judgment task remained high across all motion conditions, reaction time increased with motion intensity, particularly revealing a significant interaction with susceptibility. This pattern exemplifies a classic speed-accuracy tradeoff and provides direct evidence for the hidden cognitive cost of motion. It suggests that participants were not unable to perform the task, but had to exert more effort to do so.

This phenomenon is best explained by the Compensatory Control Model within the Cognitive-Energetical Theory \citep{hockey2011motivational}. This model posits that when faced with a stressor---in this case, physical motion---the brain's executive control system actively mobilizes additional mental effort to protect the primary task goal, which is accuracy. The adoption of this high-effort, compensatory state is not without consequence. The cost of this additional effort is paid for with time, manifesting directly as the prolonged reaction times we observed. Therefore, the increase in reaction time should not be interpreted as a passive failure of processing, but as an active, strategic, and resource-intensive act of cognitive regulation.

\subsection{The Duality of Motion}
Our analysis of the decomposed motion components was designed to answer RQ2 and RQ3, which questioned whether motion has distinct opposing effects. Our findings affirmed this duality. In response to RQ2, the analysis revealed that the Lateral Interference component was a significant positive predictor of reaction time. In response to RQ3, the analysis showed that the Directional Congruency component was a significant negative predictor of reaction time.

We can unify these two seemingly contradictory effects under the theoretical framework of Predictive Coding \citep{talsma2015predictive}. This theory frames the brain as a prediction machine that constantly generates internal models to anticipate sensory input, and works to minimize the prediction error between its predictions and the actual sensory signals.

From this perspective, the detrimental effect of Lateral Interference can be understood as a state of high and persistent prediction error. The lateral motion creates a strong sensory conflict between the visual system (which perceives a stable, task-focused world) and the vestibular system (which signals bodily motion). The brain's internal model, which predicts stability to perform the task, is constantly violated by the incoming vestibular signals. According to the predictive coding framework, resolving this large and continuous prediction error is computationally expensive, requiring significant cognitive resources that are then diverted from the primary directional judgment task \citep{talsma2015predictive}.

In stark contrast, the facilitative effect of Directional Congruency represents a state of minimized prediction error. When the physical motion aligns with the direction implied by the visual cue, the vestibular and proprioceptive inputs perfectly match the brain's prediction, which is primed by the visual stimulus. This is not merely the absence of conflict but an active state of sensorimSotor synergy. In this state, the physical motion ceases to be passive context and becomes an integral part of the cognitive computation itself, directly facilitating the perception-action loop. This finding is conceptually analogous to principles demonstrated in pseudo-haptics, where congruent visual cues can create compelling physical illusions \citep{10.3389/frvir.2023.973083}. Our work extends this principle by showing that congruent physical cues can likewise enhance cognitive processing. This result provides foundational empirical evidence that motion can be an ally to cognition, justifying a paradigm shift from merely suppressing motion to actively shaping it.

\subsection{The Asymmetric Impact of Susceptibility}

A key finding of this study addresses RQ4 and RQ5, which investigated the moderating role of individual susceptibility. We identified an asymmetry in how susceptibility moderates motion's effects: in response to RQ4, susceptibility significantly amplified the negative impact of Lateral Interference. However, in response to RQ5, susceptibility did not significantly alter the positive impact of Directional Congruency. We propose that this asymmetry stems from the differential reliance of these two processes on a specific executive function: inhibitory control.

Benefiting from congruency appears to be a relatively automatic process of multisensory integration that occurs when signals are naturally aligned, requiring minimal top-down regulation. In contrast, coping with the sensory conflict induced by Lateral Interference is an active, effortful process that requires the brain to inhibit or suppress the salient but task-irrelevant vestibular signals to prioritize the visual information. This act of suppression is a core component of inhibitory control, a key executive function. This interpretation that lateral interference is a cognitive load rather than a simple physiological one is critically supported by recent findings. For example, research on motion sickness anisotropy found that under conditions of pure physical motion without a concurrent cognitive task, lateral motion was actually less nauseogenic and induced significantly lower sickness symptoms than longitudinal motion \citep{sato2026anisotropy}. The apparent contradiction between this and our finding where lateral interference impaired performance precisely supports our hypothesis. The detriment we measured is not likely a product of motion sickness; it is the cognitive cost of inhibiting a task-orthogonal signal that competes for directional resources.

Recent neurophysiological evidence has linked visually induced motion sickness to impaired inhibitory control, as indexed by a reduction in the P3 ERP component \citep{WU2020102981}. This prior work is consistent with our behavioral asymmetry and suggests a plausible account, but we did not directly measure inhibitory control or neural markers in the present study. Accordingly, we interpret MSSQ as a susceptibility proxy that may covary with the efficiency of suppressing task-irrelevant vestibular signals. Under this account, individuals with higher susceptibility would incur a larger cognitive cost when inhibiting lateral motion, yielding disproportionate RT slowing. In contrast, because the congruency condition relies less on active suppression and more on automatic multisensory alignment, susceptibility would be expected to exert a weaker influence on its benefit. Future work should test this mechanism directly by adding an independent inhibitory-control measure or concurrent EEG indices, alongside subjective strategy reports and state measures, to identify when congruency remains facilitative versus when it becomes additional load.

\subsection{Implications for Design: From Motion Suppression to Motion Shaping}
Our findings have profound implications for the design of interactive systems in mobile environments, supplementing the traditional paradigm of Motion Suppression and advocating for a new approach of Motion Shaping. The prevailing engineering goal of simply minimizing all motion is both practically difficult and, as our results show, conceptually flawed because it ignores the potential for motion to be beneficial \citep{Metzulat2024-visualInput,Karjanto2018-pvfs,KUIPER2020103068}.

We propose that designers should treat motion not as a monolithic enemy to be eliminated, but as a design material to be sculpted. The core principle of Motion Shaping is to actively design the micro-dynamics of a vehicle or platform to minimize sensory conflict while maximizing sensorimotor synergy with the user's ongoing tasks. For instance, when a navigation system is about to display a left-turn instruction, the vehicle's active suspension could generate a subtle, sub-threshold lateral acceleration to the left. This congruent motion cue could prime the user's sensorimotor system, leveraging the universal benefit of directional congruency to make their response faster and more intuitive.

This leads to a broader vision for a new Cognitive Ergonomics for mobile HCI. Its goal is not just physical comfort, but the design of environments that are adapted to the user's dynamic cognitive state. This could involve personalized cockpits that adjust their motion-damping profiles based on a user's known susceptibility, or adaptive interfaces that shift from visual to auditory tasks when the system detects a period of high lateral interference.

This design stance resonates with somaesthetic interaction design, which advocates engaging the lived body as design material rather than treating it as disturbance. In this view, bodily sensations and micro-dynamics are resources to be shaped for meaning and performance, not noise to be eliminated \citep{hook2018-designing}. Our results operationalize this stance for mobile HCI: by minimizing geometric conflict and aligning motion with the demanded action, motion becomes part of the computation that scaffolds perception–action loops.

At the same time, cue-based mitigation should be treated as an empirical design space rather than a guaranteed fix: augmented visual environments can yield little additional benefit beyond an external view \citep{deWinkel2021-augmentedVisual}, and pre-registered comparisons of cue modalities have reported null or small effects under controlled motion exposure \citep{Reuten2024-anticipatoryCues}.

\subsection{Limitations and Future Work}
This study has several limitations that open avenues for future research. First, the experiment was conducted in a high-fidelity simulator. While this offered excellent experimental control, it may not fully replicate the multimodal complexities of real-world driving, and future work should aim to validate these findings in on-road studies. Second, our motion profile focused on low-frequency, large-amplitude swaying characteristic of vehicle dynamics, but did not include high-frequency, low-amplitude vibration. Future studies should investigate how this different type of motion, also prevalent in vehicles, affects cognitive performance. Finally, our cognitive task was a simple directional judgment task using keypresses. The generalization of these motion effects to more complex tasks, such as reading or interacting with interfaces, remains an open question. Moreover, future studies could employ a touchscreen to better simulate interactions with in-car center consoles and investigate how motion impacts fine motor tasks like targeting and selection. Our motion profile sampled lateral perturbations under controlled lab conditions and does not exhaust the full space of whole-body vibration. In particular, we did not test roll/pitch combinations, extreme magnitudes, or very low-frequency components that ISO 2631-1 associates with motion-sickness–relevant bands. Caution is therefore warranted when generalizing our results across axes and frequency bands of WBV \citep{ISO2631-1}.

This extension is not merely incremental: vehicle-like vibration has been shown to systematically impair touchscreen operations. In controlled experiments, vibration increased touch errors and task time, with especially large costs for high-precision targets; lowering precision requirements (e.g., larger targets) mitigated part of the deficit \citep{TAO2021103293}. Complementarily, comparative studies found that while touch is fastest in static conditions, under vibration its error rates surge relative to indirect devices, underscoring that what is optimal at rest may fail in motion. These findings motivate our proposed touchscreen follow-ups and suggest concrete manipulations (target size, control-to-display gain) for testing whether motion shaping can offset fine-motor degradation.

Future research could also explore novel experimental paradigms. One promising direction is to investigate dual-dynamic scenarios, where the visual target itself moves within the moving environment, to understand the interplay between these two sources of motion. Another approach could simplify the motion within trials; instead of using complex superimposed sine waves, employing discrete, directional movements (e.g., a single surge forward or sway to the left) could allow for a more direct and granular examination of the congruency effect. Furthermore, integrating our motion platform with virtual reality would enable the systematic study of the congruence and incongruence between physical motion and first-person virtual motion, offering deeper insights into their combined impact on cognitive performance and subjective experiences like cybersickness.

Lastly, our interpretation of the role of inhibitory control is based on behavioral data. Future studies employing neurophysiological measures, such as EEG, could directly test this hypothesis by examining neural markers of inhibitory control (e.g., the P3 component) during task performance under different motion conditions. Generalizability is also bounded by our sample. Participants were predominantly young adults, and while we recorded trait susceptibility, the range likely under-represents both tails of the distribution. The MSSQ provides reliable percentile norms for classic motion sickness, yet it does not capture visually induced motion sickness per se; future work should stratify recruitment by MSSQ percentiles and consider complementary VIMS-focused instruments when interfaces rely heavily on visual motion \citep{Golding2006}.

\subsection{Guidelines for HCI}
This subsection translates our findings into interface-level guidance for both drivers and passengers. The evidence indicates that vehicle motion is not uniformly detrimental. When interface cues are aligned with the direction the body already senses, responses become faster; when side-to-side disturbances dominate, responses slow, especially for users who are more susceptible to motion. Rather than treating motion as a nuisance to be suppressed entirely, the goal is to remove conflicts while turning aligned motion into a supportive part of the interface. This position complements long-standing efforts to reduce visual–manual distraction in vehicles \citep{NHTSA2013} and draws on classic results in multisensory integration, sensory conflict, mental workload, and resource theory \citep{ErnstBanks2002,ReasonBrand1975,HartStaveland1988,Wickens2002}.

First, align interface cues with the direction of ongoing movement whenever a task requires deciding where to go or which option to choose. In practice, visual highlights, navigational arrows, focus transitions, and brief vibrotactile confirmations should move in the same direction as the motion currently felt by the body. When perfect alignment is not feasible, prefer neutral or attenuated effects rather than cues that imply the opposite direction. Alignment reduces cross-modal conflict and exploits the way the nervous system combines redundant information across senses \citep{ErnstBanks2002,ReasonBrand1975}. It also conserves limited attentional resources by avoiding unnecessary competition between modalities \citep{Wickens2002}.

Second, manage periods of pronounced lateral disturbance as windows to avoid dense reading and precise pointing. During continuous cornering, side winds, or rough surfaces, defer non-critical interactions, enlarge targets, reduce parallel choices on a single screen, and prefer speech or brief auditory notifications. For drivers, maintain attention on the road and avoid pop-ups that demand fine visual–manual control in high-disturbance windows; for passengers, schedule text-heavy tasks for calmer segments. This strategy is consistent with established guidance to limit visual–manual workload while driving \citep{NHTSA2013}, with foundational evidence on workload assessment \citep{HartStaveland1988}, and with standards that characterize the discomfort and fatigue associated with lateral vibration \citep{ISO2631-1}.

Third, provide a lighter interaction path for users who are more affected by motion. Offer an optional reduced-load mode and switch to it automatically when the system detects substantial slowing or more frequent errors, prioritizing larger controls, stepwise confirmations, steadier scrolling, and low-demand modalities. For drivers, such transitions should be unobtrusive and minimize interruption; for passengers, a clear on–off control is acceptable. This recommendation follows long-standing observations that susceptibility to motion varies substantially across individuals \citep{Golding2006,ReasonBrand1975}.

Fourth, treat touchscreens conservatively under vibration. When motion intensity or lateral disturbance is detected, prefer larger targets, reduce parallel choices, and defer precision-demanding gestures; where feasible, adapt control-to-display gain to stabilize pointing. Empirical work shows that vibration disproportionately harms fine-precision touch and raises workload, while larger targets and moderated gain can recover performance \citep{TAO2021103293}.

In summary, effective motion-aware interfaces do not seek to eliminate motion altogether. They first sidestep lateral-disturbance windows and then use alignment to turn motion into a resource for guidance. This shifts the design stance from blanket suppression toward shaping at the user-interface level, improving performance without increasing distraction.

\section{Conclusion}

In conclusion, this research addresses how physical motion impacts cognitive performance in mobile HCI. Instead of treating motion as a single variable of intensity, we proposed and tested a method of decomposing it into opposing geometric components: task-irrelevant Lateral Interference and task-aligned Directional Congruency. Our experimental result demonstrates three key findings: First, an increase in lateral interference significantly lengthened participants' reaction time in visual judgment task; Conversely, an increase in directional congruency significantly amplified for individuals with high motion sickness susceptibility, whereas the beneficial effect of congruency was consistent for all participants. These results indicate that motion's impact is not monolithic, suggesting that it can be proactively shaped by minimizing specific interference components while maximizing congruency to better support cognitive performance in mobile environments.


\appendix
\section{Supplementary Descriptive Statistics}
\label{sec:app_desc_stats}

To support reproducibility and to facilitate checks for speed--accuracy trade-offs, we report participant-level descriptive statistics for accuracy and reaction time (RT) in each motion phase. These tables are provided for auditability rather than for inferential claims; inferential tests are reported in the main Results section.

\begin{table}[!t]
\centering
\caption{Supplementary descriptive statistics of accuracy by motion phase. Accuracy is the participant-level mean proportion correct within each phase. Values are reported as Mean $\pm$ SD across participants. $N$ denotes the number of participants contributing valid trials in that phase.}
\label{tab:app_acc_desc}
\begin{tabular}{lrrr}
\toprule
Condition & $N$ & Mean & SD \\
\midrule
Baseline                  & 36 & 0.997168 & 0.006772 \\
Low                       & 36 & 0.999201 & 0.003343 \\
Medium                    & 38 & 0.999243 & 0.003256 \\
High                      & 36 & 0.999195 & 0.003367 \\
Post-motion               & 37 & 0.997570 & 0.007371 \\
\bottomrule
\end{tabular}
\end{table}

\begin{table}[!t]
\centering
\caption{Supplementary descriptive statistics of reaction time (ms) by motion phase. RT is the participant-level mean across correct trials (after standard RT filtering). Values are reported as Mean $\pm$ SD across participants. $N$ denotes the number of participants contributing valid data in that phase.}
\label{tab:app_rt_desc}
\begin{tabular}{lrrr}
\toprule
Condition & $N$ & Mean (ms) & SD (ms) \\
\midrule
Baseline                  & 36 & 513.714 &  76.484 \\
Low                       & 36 & 548.147 & 138.301 \\
Medium                    & 38 & 524.545 &  90.992 \\
High                      & 36 & 535.866 & 103.504 \\
Post-motion               & 37 & 530.517 & 119.864 \\
\bottomrule
\end{tabular}
\end{table}

\section{Post-hoc pairwise contrasts by MSSQ probe values}
\begin{sidewaystable*}[p]
\centering
\caption{Tukey-adjusted pairwise contrasts of motion conditions at three probe values of MSSQ (scaled: $-1$ $\mathit{SD}$, mean, $+1$ $\mathit{SD}$). $\hat{\beta}$ is the contrast on the log(seconds) scale; ratio $=\exp(\hat{\beta})$ indicates multiplicative change in reaction time; \% change $=(\text{ratio}-1)\times 100$. Confidence intervals are Wald 95\% intervals as returned by \texttt{emmeans}.}
\label{tab:posthoc_pairs_by_mssq}
\scriptsize
\setlength{\tabcolsep}{4pt}
\renewcommand{\arraystretch}{1.15}
\begin{tabular}{llrrrrrllrl}
\toprule
MSSQ & Contrast (A vs B) & \textit{estimate} & SE & $z$ & $p_{\mathrm{Tukey}}$ & ratio & 95\% CI (ratio) & \% change & 95\% CI (\%) \\
\midrule
$-1$ SD & Baseline vs Low & -0.008 & 0.007 & -1.16 & .774 & 0.992 & [0.973, 1.011] & -0.8 & [-2.7, 1.1] \\
$-1$ SD & Baseline vs Medium & 0.007 & 0.007 & 0.96 & .873 & 1.007 & [0.988, 1.026] & 0.7 & [-1.2, 2.6] \\
$-1$ SD & Baseline vs High & 0.007 & 0.007 & 1.04 & .836 & 1.007 & [0.988, 1.027] & 0.7 & [-1.2, 2.7] \\
$-1$ SD & Baseline vs Post-motion & 0.015 & 0.007 & 2.15 & .197 & 1.015 & [0.996, 1.035] & 1.5 & [-0.4, 3.5] \\
$-1$ SD & Low vs Medium & 0.015 & 0.007 & 2.13 & .206 & 1.015 & [0.996, 1.035] & 1.5 & [-0.4, 3.5] \\
$-1$ SD & Low vs High & 0.016 & 0.007 & 2.21 & .176 & 1.016 & [0.996, 1.035] & 1.6 & [-0.4, 3.5] \\
$-1$ SD & Low vs Post-motion & 0.023 & 0.007 & 3.32 & .008 & 1.024 & [1.004, 1.044] & 2.4 & [0.4, 4.4] \\
$-1$ SD & Medium vs High & 0.001 & 0.007 & 0.10 & .999 & 1.001 & [0.982, 1.020] & 0.1 & [-1.8, 2.0] \\
$-1$ SD & Medium vs Post-motion & 0.008 & 0.007 & 1.19 & .739 & 1.008 & [0.989, 1.027] & 0.8 & [-1.1, 2.7] \\
$-1$ SD & High vs Post-motion & 0.008 & 0.007 & 1.11 & .790 & 1.008 & [0.989, 1.027] & 0.8 & [-1.1, 2.8] \\
Mean & Baseline vs Low & -0.027 & 0.005 & -5.34 & \textless.001 & 0.973 & [0.960, 0.987] & -2.7 & [-4.0, -1.3] \\
Mean & Baseline vs Medium & -0.024 & 0.005 & -4.85 & \textless.001 & 0.976 & [0.963, 0.990] & -2.4 & [-3.7, -1.0] \\
Mean & Baseline vs High & -0.029 & 0.005 & -5.79 & \textless.001 & 0.971 & [0.958, 0.985] & -2.9 & [-4.2, -1.5] \\
Mean & Baseline vs Post-motion & -0.006 & 0.005 & -1.18 & .744 & 0.994 & [0.981, 1.007] & -0.6 & [-1.9, 0.7] \\
Mean & Low vs Medium & 0.003 & 0.005 & 0.62 & .971 & 1.003 & [0.990, 1.017] & 0.3 & [-1.0, 1.7] \\
Mean & Low vs High & -0.002 & 0.005 & -0.37 & .995 & 0.998 & [0.985, 1.012] & -0.2 & [-1.5, 1.2] \\
Mean & Low vs Post-motion & 0.021 & 0.005 & 4.15 & \textless.001 & 1.021 & [1.007, 1.035] & 2.1 & [0.7, 3.5] \\
Mean & Medium vs High & -0.005 & 0.005 & -0.99 & .853 & 0.995 & [0.982, 1.009] & -0.5 & [-1.8, 0.9] \\
Mean & Medium vs Post-motion & 0.018 & 0.005 & 3.60 & .003 & 1.018 & [1.004, 1.032] & 1.8 & [0.4, 3.2] \\
Mean & High vs Post-motion & 0.023 & 0.005 & 4.60 & \textless.001 & 1.023 & [1.009, 1.037] & 2.3 & [0.9, 3.7] \\
$+1$ SD & Baseline vs Low & -0.046 & 0.007 & -6.43 & \textless.001 & 0.955 & [0.937, 0.974] & -4.5 & [-6.3, -2.6] \\
$+1$ SD & Baseline vs Medium & -0.055 & 0.007 & -8.14 & \textless.001 & 0.947 & [0.930, 0.964] & -5.3 & [-7.0, -3.6] \\
$+1$ SD & Baseline vs High & -0.066 & 0.007 & -9.14 & \textless.001 & 0.936 & [0.918, 0.955] & -6.4 & [-8.2, -4.5] \\
$+1$ SD & Baseline vs Post-motion & -0.027 & 0.007 & -3.88 & .001 & 0.973 & [0.954, 0.992] & -2.7 & [-4.6, -0.8] \\
$+1$ SD & Low vs Medium & -0.009 & 0.007 & -1.29 & .668 & 0.992 & [0.973, 1.012] & -0.9 & [-2.7, 1.2] \\
$+1$ SD & Low vs High & -0.020 & 0.007 & -2.80 & .054 & 0.980 & [0.960, 1.000] & -2.0 & [-4.0, 0.0] \\
$+1$ SD & Low vs Post-motion & 0.018 & 0.007 & 2.51 & .111 & 1.018 & [0.998, 1.038] & 1.8 & [-0.2, 3.8] \\
$+1$ SD & Medium vs High & -0.011 & 0.007 & -1.54 & .525 & 0.989 & [0.970, 1.009] & -1.1 & [-3.0, 0.9] \\
$+1$ SD & Medium vs Post-motion & 0.027 & 0.007 & 3.83 & .001 & 1.027 & [1.008, 1.047] & 2.7 & [0.8, 4.7] \\
$+1$ SD & High vs Post-motion & 0.038 & 0.007 & 5.21 & \textless.001 & 1.039 & [1.018, 1.060] & 3.9 & [1.8, 6.0] \\
\bottomrule
\end{tabular}
\end{sidewaystable*}

\section{Supplementary distribution diagnostics}
\label{app:diagnostics}
\begin{figure}[!h]
    \includegraphics[width=\textwidth]{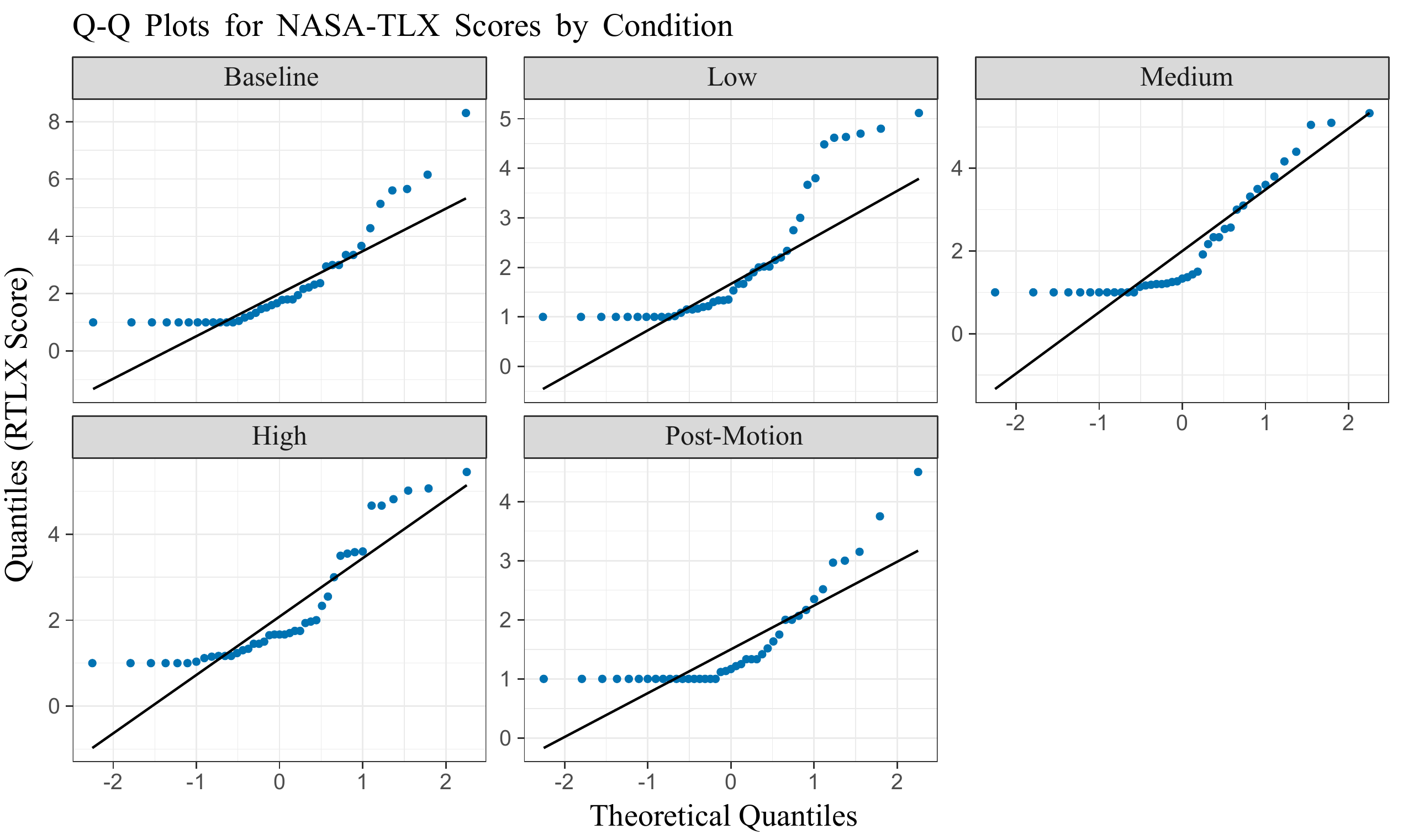}
    \caption{Supplementary Q--Q plots for NASA-TLX (RTLX) total scores across the five experimental phases, used to assess departures from normality prior to selecting non-parametric analyses.}
    \label{fig:nasa_qq}
\end{figure}

\bibliographystyle{elsarticle-harv}
\bibliography{references}

\end{document}